\documentclass{aa} 
\usepackage{graphicx}
\usepackage{txfonts}
\usepackage{natbib}

\begin{document}
\title{Simultaneous X-ray, radio, near-infrared, and optical monitoring of Young Stellar Objects in the \textsl{\textbf{Coronet}} cluster}
\author{J. Forbrich\inst{1,2} \and Th. Preibisch\inst{1} \and K. M. Menten\inst{1} \and R. Neuh\"auser\inst{2} \and F. M. Walter\inst{3} \and M. Tamura\inst{4} \and N. Matsunaga\inst{5} \and N.~Kusakabe\inst{6} \and Y. Nakajima\inst{4} \and A. Brandeker\inst{7} \and S. Fornasier\inst{8} \and B. Posselt\inst{9,2} \and K. Tachihara\inst{10} \and C. Broeg\inst{2}}
\offprints{Jan Forbrich, \email{forbrich@mpifr-bonn.mpg.de}}
\institute{
Max-Planck-Institut f\"ur Radioastronomie, Auf dem H\"ugel 69, D-53121 Bonn, Germany
\and
Astrophysikalisches Institut und Universit\"ats-Sternwarte Jena, Schillerg\"a{\ss}chen 2-3, D-07745 Jena, Germany
\and 
Department of Physics and Astronomy, Stony Brook University, Stony Brook, NY 11794-3800, U.S.A.
\and
National Astronomical Observatory of Japan, Mitaka, Tokyo 181-8588, Japan
%Motohide Tamura
\and
Institute of Astronomy, School of Science, The University of Tokyo, Osawa 2-21-1, Mitaka, Tokyo 181-0015, Japan
%Noriyuki Matsunaga
\and
Department of Astronomical Sciences, Graduate University for Advanced Studies (Sokendai), Mitaka, Tokyo 181-8588, Japan
%Nobuhiko Kusakabe
\and
Department of Astronomy and Astrophysics, University of Toronto, 60 St. George Street, Toronto, ONT M5S 3H8, Canada
\and
%Dipartimento di Astronomia and CISAS, Universit\`a di Padova, Vicolo dell'Osservatorio 5, 35100 Padova, Italy
Universit\'e Paris 7, LESIA -- Observatoire de Paris, B\^atiment 17bis, 5 Place Jules Janssen, F-92195 Meudon Principal Cedex, France
\and
Max-Planck-Institut f\"ur extraterrestrische Physik, Giessenbachstr. 1, D-85748 Garching, Germany
\and
Graduate School of Science and Technology, Kobe University, 1-1 Rokko-dai Nada-ku, Kobe 657-8501, Japan
%Kengo
}
 
\date{Received; accepted}

\abstract{Multi-wavelength (X-ray to radio) monitoring of Young Stellar Objects (YSOs) can provide important information about physical processes at the stellar surface, in the stellar corona, and/or in the inner circumstellar disk regions. While coronal processes should mainly cause variations in the X-ray and radio bands, accretion processes may be traced by time-correlated variability in the X-ray and optical/infrared bands. Several multi-wavelength studies have been successfully performed for field stars and $\sim 1-10$~Myr old T Tauri stars, but so far no such study succeeded in detecting simultaneous X-ray to radio variability in extremely young objects like class~I and class~0 protostars.}{Here we present the first simultaneous X-ray, radio, near-infrared, and optical monitoring of YSOs, targeting the \textsl{Coronet} cluster in the Corona Australis star-forming region, which harbors at least one class~0 protostar, several class~I objects, numerous T~Tauri stars, and a few Herbig AeBe stars.}{In August 2005, we obtained five epochs of \textsl{Chandra} X-ray observations on nearly successive days accompanied by simultaneous radio observations at the NRAO Very Large Array during four epochs, as well as by simultaneous optical and near-infrared observations from ground-based telescopes in Chile and South Africa.}{Seven objects are detected simultaneously in the X-ray, radio, and optical/infrared bands; they constitute our core sample. While most of these sources exhibit clear variability in the X-ray regime and several also display optical/infrared variability, none of them shows significant radio variability on the timescales probed. We also do not find any case of clearly time-correlated optical/infrared and X-ray variability. Remarkable intra-band variability is found for the class~I protostar IRS~5 which shows much lower radio fluxes than in previous observations, and the Herbig Ae star R~CrA, which displays enhanced X-ray emission during the last two epochs, but no time-correlated variations are seen for these objects in the other bands. The two components of S~ CrA vary nearly synchronously in the $I$ band.}{The absence of time-correlated multi-wavelength variability suggests that there is no direct link between the X-ray and optical/infrared emission and supports the notion that accretion is {{\it not}} an important source for the X-ray emission of these YSOs. No significant radio variability was found on timescales of days.}


\titlerunning{Simultaneous multi-wavelength observations of YSOs in the \textsl{Coronet} cluster}

\maketitle

\section{Introduction}

Before low-mass young stars reach the main sequence they mature through a series of evolutionary stages termed class~0 to class~III (\citealp{lad87}, and e.g. \citealp{and00}). Named 'Young Stellar Objects' (YSOs) throughout this development, in the earliest stages they are referred to as 'protostars'. A class~0 protostar, observed as cold submillimeter source still has to accrete most of its mass. Very soon a disk and a bipolar outflow develop. The spectral energy distribution slowly shifts towards shorter wavelengths: A class~I protostar is already seen in the near-infrared even though its luminosity still is mostly due to accretion (at a canonical age of about $10^5$ years). In the subsequent classical and weak-line T Tauri stages (class~II and III), most of the material has already been accreted leaving less and less circumstellar material. Flaring in YSOs is a ubiquitous phenomenon at diverse wavelengths, however it has not previously been studied simultaneously across the electromagnetic spectrum in protostars and rarely in more evolved YSOs, even though multi-wavelength correlations would contain additional information on underlying physics.

One would expect some correlation of thermal hot-plasma X-ray emission and non-thermal centimetric radio emission interpreted as gyrosynchrotron radiation due to magnetic fields because both types of radiation are partially produced in the same regions. For a discussion of radio and X-ray observations of (proto-)stellar coronae, see \citet{gue02,gue04}, and for a discussion of multi-wavelength aspects, see \citet{gue02mw}. Looking at the sun as the star which is best studied for multi-wavelength variability, we see much uncorrelated variability but also a particular kind of radio--X-ray correlation in the form of the Neupert effect which was observed towards a star other than the sun (the dMe flare star UV Cet) for the first time by \citet{gue96}.  Radio and X-ray emission both trace the innermost surroundings of protostars at less than a few stellar radii while near-infrared emission is mainly due to warm dust. 

A correlation of optical/near-infrared flares interpreted as accretion events and X-ray variability would permit speculations on the connection of accretion and X-ray emission. Such a correlation can be expected from the observation of different X-ray luminosities for accreting and non-accreting T~Tauri stars as well as a few cases where soft X-ray emission was interpreted as due to accretion (see Section \S \ref{rxsec}). Another cause for correlated variability is large cool star spots due to magnetic activity. They lead to rotational modulation of photospheric light and may directly or indirectly influence X-ray emission. In the near-infrared, variations in the circumstellar disk emission are more important than variability in the stellar photosphere \citep{eir02}. In case of X-ray emission due to accretion shocks (see next subsection), one would expect to see a correlation at near-infrared or even optical wavelengths, depending on the accretion rate. For the earliest evolutionary stages of stars, a direct comparison of optical and near-infrared variability is not possible since they can only be observed in the near-infrared if at all. Variability at near-infrared and optical wavelengths can also be due to pulsation and eclipsing multiple systems.

\begin{figure*}
\begin{minipage}{12cm}
 \includegraphics[width=12cm]{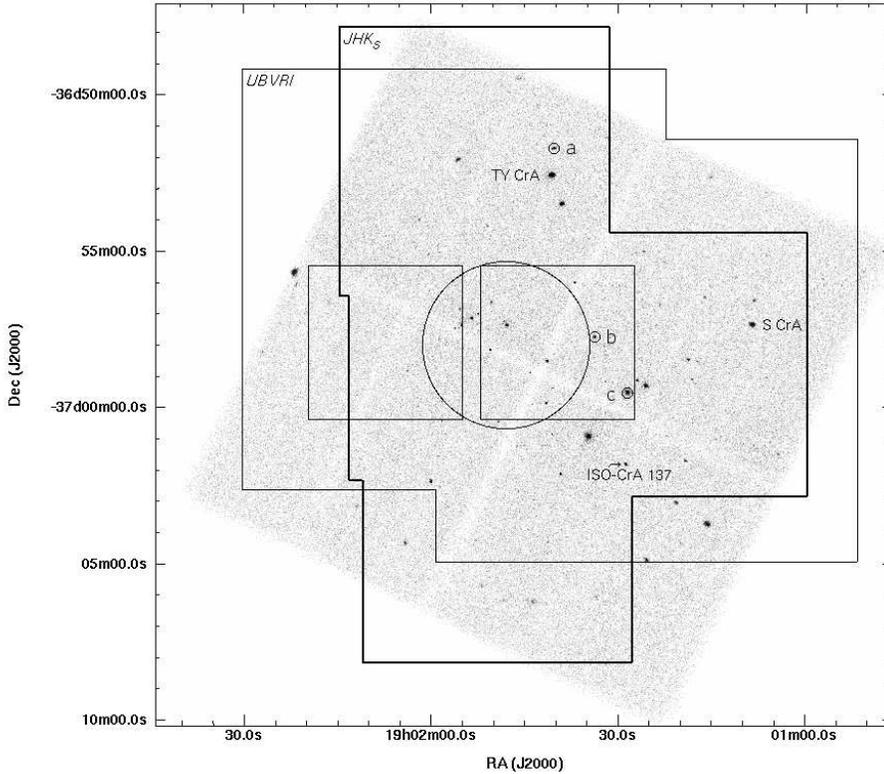}
\end{minipage}
\begin{minipage}{6cm}
 \vspace*{4.7cm}
 \caption{Plot of the \textsl{Coronet} region covered in this multi-wavelength dataset. The background image shows the combined X-ray data, the big circle indicates the half-power beam width of the VLA primary beam. The outer two boxes mark the coverage of the SMARTS-\textsl{UBVRI} and SIRIUS-$JHK_S$ data (thick line). The two small boxes show the fields of view of the SofI observations. The central \textsl{Coronet} cluster is shown in detail in Fig.~\ref{vlacxo}. Three 2MASS sources discussed in the text are marked as small circles: a) 2MASS~J19014041--3651422 b) 2MASS~J19013385--3657448 c) 2MASS~J19012872--3659317}
 \end{minipage}
\label{mwplot}
\end{figure*}

\subsection{Radio and X-ray emission from YSOs}
\label{rxsec}

YSOs are known to emit thermal as well as non-thermal centimetric radio emission. While the thermal emission appears to be mostly due to shock-induced ionisation at the base of outflows, the non-thermal emission mainly is gyrosynchrotron emission from mildly relativistic electrons gyrating in magnetic fields. For a review of radio properties of YSOs, see \citet{and96}; a review of associated high-energy processes was given by \citet{fem99}.

The main possibilities for the origin of the strong X-ray radiation from
YSOs are (1) emission from  a magnetically confined plasma 
(either in stellar coronae similar to that of the Sun and other late-type stars, or perhaps in magnetic structures connecting the star and its circumstellar disk) with magnetic reconnection causing flares and (2) emission from hot accretion shocks at the stellar surface. Recent sensitive X-ray observations of large samples of T Tauri stars (e.g. the \textsl{Chandra} Orion Ultradeep Project; see \citealp{get05}) have now clearly established that the bulk of the X-ray emission from T Tauri stars must arise from magnetically confined plasma, most likely in rather compact coronal loops. Accretion shocks seem to contribute some fraction of the soft X-ray emission of \textsl{some} TTS \citep[e.g.~TW Hya, XZ Tau and BP Tau, see][]{kas02,sch05}, but clearly cannot account for the high X-ray luminosities and plasma temperatures seen in the majority of TTS \citep{pre05}.
Nevertheless, there is a relation between X-ray activity and disk accretion in the sense that actively accreting TTS are \textsl{weaker} X-ray emitters than non-accreting TTS. This effect may be even more important for class~I protostars. Possible explanations for this suppression of X-ray emission in accreting TTS are discussed in \citet{pre05} and in \citet{jar06}.

X-ray emission from class~I protostars is also well established observationally \citep{ikt01,pre04}, and an X-ray flare towards a candidate class~0 protostar in CrA was reported by \citet{ham05}. The very high plasma temperatures derived from the protostellar X-ray spectra again clearly suggest an origin related to some kind
of magnetic activity, whereas accretion processes cannot reproduce the
observed X-ray properties.

Until now, there have been few simultaneous multi-wavelength observations of YSOs, mainly targeting T Tauri stars and not class~0 or class~I protostars: \citet{fei94} observed a magnetically active T Tauri star (V773~Tau) simultaneously at  X-ray, ultraviolet, optical, and radio wavelengths, while \citet{gul97} and \citet{ste03} study simultaneous optical and X-ray variability of the classical T Tauri star BP Tau and the weak-line T Tauri star V410 Tau, respectively. \citet{gue00} targeted pre-main sequence stars in the Taurus-Auriga region, specifically V773~Tau, in quasi-simultaneous X-ray, optical, and radio observations. In all cases, no correlated variability was found, based on relatively short observations. \citet{bow03} report the simultaneous radio and X-ray observation of an enormous radio flare towards a supposed highly obscured weak-line T Tauri star (GMR-A) in Orion. The X-ray flux increased by a factor of 8 two days before the radio flare detection. The $\rho$ Oph cloud complex and its pre-main sequence stars were selected for simultaneous radio and X-ray observations by \citet{gag04}, covering the largest number of sources yet. However, none of the class~I protostars were detected in both wavelength regimes. \citet{ber05} attempted to constrain the magnetic properties of a candidate brown dwarf by simultaneous radio, X-ray, and H$\alpha$ observations. Because of the low upper limit to the X-ray flux, they found a large radio-to-X-ray luminosity ratio. Most recently, \citet{sta06} present a comprehensive optical variability study of Orion pre-main sequence stars, carried out simultaneously with the \textsl{Chandra} Orion Ultra-Deep Project (COUP). They find indications for time-correlated optical and X-ray variability in only 5\% of their sources and conclude that there is little evidence for a causal link between variability at optical and X-ray wavelengths.

In a preparatory study to this multi-wavelength project, we singled out the Corona Australis star-forming region, particularly the \textsl{Coronet} cluster, as the ideal target for this kind of observation: It is a compact cluster of protostars and other YSOs, most of them detected at radio, infrared and X-ray wavelengths (\citealp{for05}; for a review of the entire Corona Australis star-forming region, see \citealp{nef06}).

\section{Observations and data reduction}

\begin{figure}[h]
\includegraphics*[width=8.7cm,bb= 25 385 590 720]{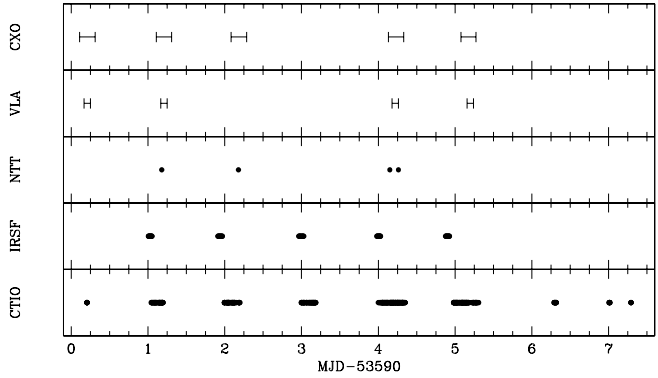}
\begin{center}
\caption{Graphical depiction of the observing times at the different observatories indicated on the lefthand side.}
\label{organizer}
\end{center}
\end{figure}

We present here the results of a multi-wavelength observing campaign targeting YSOs in Corona Australis, based on five epochs of \textsl{Chandra} observations. For the first time, protostars (class~0/I YSOs) were observed simultaneously at radio, X-ray, and near-infrared wavelengths. The field of view of the observations is shown in Fig.~\ref{mwplot}, their scheduling is shown in Fig.~\ref{organizer} as well as listed in Table~\ref{obslog}.

Due to the far southern declination of the \textsl{Coronet} cluster (nearly $-37^\circ$), the region is accessible for the NRAO\footnote{The National Radio Astronomy Observatory is a facility of the National Science Foundation operated under cooperative agreement by Associated Universities, Inc.} Very Large Array only during a few hours per day. Thus, the core experiment was designed around five 4h duration \textsl{Chandra} observations on successive days (with a one-day gap due to perigee passage when the satellite is inoperative). The instrument used was the Advanced CCD Imaging Spectrometer (ACIS). The \textsl{Chandra} observations were to be complemented by five epochs of simultaneous VLA observations around the \textsl{Coronet} Cluster's meridian passage. To ease scheduling, blocks of 2h VLA time were to be scheduled within the 4h \textsl{Chandra} observations. 
A few simultaneous near-infrared observations were collected with ESO's 3.5m New Technology Telescope (NTT) in Chile, using the SofI imager in the $JHK_S$ bands subsequently and in two fields east and west of R~CrA. SofI is based on a 1024x1024 Hawaii HgCdTe CCD detector, providing a field of view of $4\farcm94\times4\farcm94$ with a pixel scale of $0\,\farcs288$ (using the Large Field Objective). The exposure time was set to the minimal possible value at 1.182~seconds.

Additional near-simultaneous observations were carried out with SIRIUS (Simultaneous Infrared Imager for Unbiased Survey) at the 1.4m telescope of the InfraRed Survey Facility\footnote{IRSF is  operated by Nagoya University and SAAO (South African Astronomical Observatory) at Sutherland, South Africa.} (IRSF) in South Africa (for a description of this simultaneous three-band NIR camera, see \citealp{nag03}). The SIRIUS camera has a field of view of $7\farcm7\times7\farcm7$ and a pixel scale of $0\,\farcs45$. Four different fields were observed, one centered on the \textsl{Coronet} cluster, and three more north, west, and south of the central field. The SIRIUS images typically consist of ten dithered exposures of 30~sec duration.

Lastly, simultaneous optical imaging (bands $UBVRI$ successively) of the region was performed with the CTIO 0.9m telescope, operated by the SMARTS consortium, at the Cerro Tololo Inter-American Observatory in Chile. The detector used was a 2048x2046 CCD camera providing a field of view of $13\farcm5\times13\farcm5$ with a pixel scale of $0\,\farcs396$. Again, two fields were observed, east and west of R~CrA, but both overlapping to cover R~CrA. The last three epochs including the last multi-wavelength epoch were increasingly compromised by bad weather. In all filters and both fields, long and short exposures were taken, e.g. for the $I$-band typically durations of 20~sec and 2~sec.

\begin{table}
%\begin{center}
\caption[]{Simultaneous observations of the \textsl{Coronet} with \textsl{Chandra} and the VLA (given in UT)}
\begin{tabular}{llll}
\hline
\hline
Day (MJD$^1$)           & \textsl{Chandra}     & VLA          \\ %& ESO$^1$ & IRSF$^2$    & CTIO$^3$\\
\hline
Aug 08, 2005  (0) & 02:37:52 -- 07:33:11 & 04:04:20 -- 06:03:45 \\ %& --  & --  & yes\\
Aug 09, 2005  (1) & 02:38:52 -- 07:22:10 & 04:00:50 -- 06:00:00 \\ %& yes & yes & yes\\
Aug 10, 2005  (2) & 01:58:25 -- 06:53:28 & --		      \\ %& yes & yes & yes\\
%Aug 11, 2005  (3) & --		       & --		      \\ %& --  & yes & yes\\
Aug 12, 2005  (4) & 03:12:57 -- 07:54:57 & 04:18:30 -- 06:18:20 \\ %& yes & yes & yes\\
Aug 13, 2005  (5) & 01:51:20 -- 06:34:32 & 03:44:40 -- 05:44:10 \\ %& --  & --  & yes\\
\hline
\label{obslog}
\end{tabular}

$^1$ MJD-53590
\end{table}

\subsection{X-ray observations}

\begin{figure}
      \includegraphics*[width=8.8cm, angle=-90]{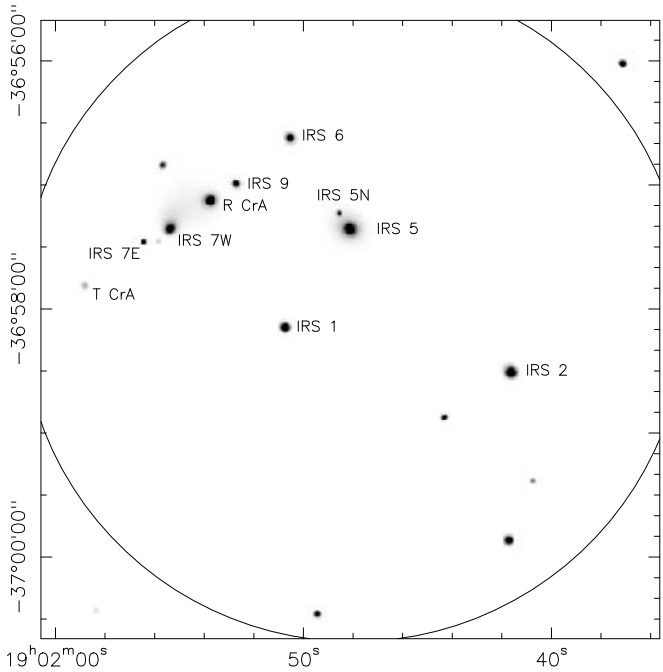}

\begin{center}
\caption{Combined \textsl{Chandra} data (Obs ID 5402-5406, smoothed). The circle denotes the approximate VLA half-power primary beam. }
\label{vlacxo}
\end{center}
\end{figure}

The \textsl{Chandra} Interactive Analysis of Observations (CIAO) 3.3 software package together with CALDB 3.2.1 was used for the analysis of the \textsl{Chandra}-ACIS data.
Apertures of $5''$ in diameter were used for light curves and spectra alike. The spectra were compiled with the task \textsl{psextract} which includes the creation of response matrices and ancillary response files. Spectral bin sizes were set to ensure a minimum of 15 events per bin.

\subsection{Radio observations}

The VLA was observing in its D configuration (program ID S6 255) at $\nu=8.44$~GHz with two intermediate frequency (IF) pairs, offset by the IF bandwidth of 50 MHz. One IF of each pair detected right circular polarization (RCP), the other one left circular polarization (LCP). The phase center was $(\alpha, \delta)_{ {\rm J}2000}= 19^{\rm h}$01$^{\rm m}$48$^{\rm s},-36^\circ58'00''$. The primary beam of the VLA antennas limits the field of view to approximately five arcminutes FWHM around the phase center position (Fig.~\ref{vlacxo}). Compared to the other observations reported here, the radio observations cover the smallest field, centered on the \textsl{Coronet} cluster. Data were analyzed using the NRAO Astronomical
Image Processing System (AIPS). Observations of 3C286, once per observing epoch, were used to establish an absolute flux density scale. The nearby phase calibrator 1924$-$292 was observed every seven minutes. We estimate that our absolute flux density scale thus derived is accurate to $\sim 15$~\%.

Due to the low elevation ($<19^\circ$), there are problems with phase instabilities caused by the atmosphere. After initial flagging of obviously faulty uv data, data were phase-only self-calibrated before imaging in Stokes-$I$ and Stokes-$V$ could be carried out for different time intervals.
Source positions and fluxes were determined by fitting Gaussians to each detection, using the AIPS tasks SAD and JMFIT. Splitting the single observing epochs into intervals of approximately one hour, the root mean square (rms) noise level in the Stokes $I$ and $V$ maps is $\approx 40$~$\mu$Jy; for the entire uv data this value reduces to 17~$\mu$Jy. The synthesized beam size for the combined four epochs is $6\,\farcs7\times 2\,\farcs0$ at position angle $1\fdg9$.

\subsection{Near-infrared and optical observations, photometry}

The SIRIUS-$JHK_S$ data were processed using a reduction pipeline developed as an external package of the IRAF\footnote{IRAF is distributed by the National Optical Astronomy Observatories, which is operated by the Association of Universities for Research in Astronomy, Inc. (AURA) under cooperative agreement with the National Science Foundation.} software package. Processing includes dark frame subtraction, flat-fielding, sky subtraction and finally combination of the dithered single images. 

The SofI-$JHK_S$ images, consisting of pairs of offset images, were processed using Eclipse 5.0.0. Dark frame subtraction, flat-fielding, and sky subtraction were performed.

The optical images in $UBVRI$ bands were dark-subtracted and flat-fielded using IRAF. Slightly different properties of the four CCD chips in the array had to be accounted for using the quadproc procedures. The observations were partly affected by bad weather in the last three epochs.

By correlating the near-infrared and optical data with 2MASS positions, an overall astrometric accuracy of $<0\,\farcs1$ was achieved.

Differential photometry was carried out for optical and near-infrared counterparts of X-ray sources using an algorithm described by \citet{bro05}. This procedure uses aperture photometry determined by the IRAF daophot package. For the aperture radii we chose values of around three times the radius of the point-spread function while the background was determined from surrounding annuli. The respective comparison stars are weighted down according to their variability as represented by their standard deviations, resulting in an optimum artificial comparison star. As an estimate of the errors in the resulting light curves, a linear model of the instrumental errors is self-consistently adjusted so that it reproduces the standard deviations of the least variable comparison stars before it is applied to the science source. For the sources analyzed here, measurements with resulting 1$\sigma$ errors of more than 0.05~mag were discarded. Due to the different selections of comparison stars, the zero levels of the differential magnitude scales generally are not the same. Thus, the light curves are plotted with the respective median magnitudes subtracted. Sources observed outside the respective linear detector ranges were not analyzed photometrically.

\subsubsection{S CrA}

We undertook a separate analysis of the S CrA optical photometry in the context of our long term monitoring, which consists of over 200 observations between 2003 April and 2006 June, mostly obtained with the CTIO/SMARTS 1.3m ANDICAM dual-channel photometer. This detector is described in, for example, \citet{wal04}.
Out of this long-term monitoring, we include seven ANDICAM data points in our analysis here, obtained in the nights of 2005 August 14-17. We do not make use of the short exposures because the data quality is poorer, and those images do not extend the temporal coverage.

S CrA is spatially resolvable under the typical observing conditions at CTIO. We measure the brightness of the individual components by fitting two Gaussians to the image. The separation and position angle are well-known, and are not free to vary. We constrain the two images to have the same point-spread function (PSF). The relative brightnesses of the two stars are then the relative amplitudes of the two Gaussians. 

The seeing (measured as the FWHM of the Gaussian PSF) in the $I$ band varied from $0\,\farcs7$ to $1\,\farcs4$, with a median value of $0\,\farcs9$. The seeing was stable for the first 4 nights, with a variance of $0\,\farcs06$, but went up in the remainder of the run, with the variance in the mean seeing ballooning to $0\,\farcs2$. There are no obvious seeing-dependent effects on the photometry.

We calibrate the brightness of the system in a three-step process. First, we determine the long-term brightness variations of the S CrA binary through differential photometry with respect to 9 other stars in the sparse field. We use an 8\arcsec aperture to enclose all the light from the S CrA binary. We assume that the weighted mean brightness of these nine stars does not vary significantly with time. The comparison brightness is heavily skewed to the flux of VSST 14, a pre-main sequence star, but not a known variable star, which is by far the brightest of the comparison stars in the field. To check whether our assumption of constancy is valid, we also compared the differential brightness against four other fainter stars singly. All yield the same long-term trends, suggesting that VSST~14 is not nearly as variable as S CrA and the assumption of constancy is warranted. 

We then use the mean long-term magnitude of S~CrA ($I$=9.81) from the Van 
Vleck Observatory T Tauri database\footnote{http://www.astro.wesleyan.edu/\textasciitilde bill/research/ttauri.html} to establish the
mean I-band magnitude. We assume that the mean differential magnitude in our photometric monitoring program corresponds to the mean long-term magnitude. We determine this offset and apply it to all the differential measurements. Finally, we divide the light from S CrA between the two components based on
the Gaussian fits.

\section{Single-wavelength variability}

After a discussion of the results by wavelength, we discuss multi-wavelength aspects for individual sources in section \ref{singlesource}.

\begin{table}
%\begin{center}
\caption[]{Radio flux densities of the \textsl{Coronet} sources}
\begin{tabular}{lll}
\hline
\hline
%aus sadfinal.dat, basierend auf corosc65.icl.001
%aktualisiert mit jmfit_corosc65.txt, 7.6.06
Source ID        & 2005 flux density$^1$  & 1998 flux density$^2$\\
                 & [mJy]              & [mJy]                \\
\hline
IRS 2            & $0.49 \pm 0.02$    & 0.29--0.36           \\
Source 5         & $1.14 \pm 0.02$    & 0.40--1.21           \\
IRS 5N           & $0.06 \pm 0.02$    & 0.10$^3$             \\
IRS 5            & $0.68 \pm 0.02$    & 0.75--3.28           \\
IRS 6            & $0.07 \pm 0.02$    & 0.15                 \\
IRS 1            & $0.42 \pm 0.02$    & 0.36--0.59           \\
R CrA            & $0.30 \pm 0.02$    & 0.16--0.26           \\
%Source 9         & $1.61 \pm 0.02$    & 0.86                 \\
IRS 7W$^4$       & $5.33 \pm 0.02$    & 4.11--5.29           \\
IRS 7E           & $1.77 \pm 0.02$    & 1.38--2.10           \\
\hline
\label{vlaresults}
\end{tabular}

$^1$ quoted errors are from Gaussian fitting. Our flux scale has an absolute uncertainty of $\sim 15$~\% \\
$^2$ minima and maxima (if source detected in single epochs), see \citet{for05} \\
$^3$ note that this source is not called IRS~5N in \citet{for05} but is no. 5 in the tables \\
$^4$ blended with Source 9 \\
\end{table}

\begin{table}
%siehe count rates.ods
%\begin{center}
\caption[]{X-ray count rates for the \textsl{Coronet} sources, from the five \textsl{Chandra} observations, 0.5-10~keV, corrected for effective exposure}
\begin{tabular}{lrrrrr}
\hline
\hline
Source ID        & S1        & S2        & S3        & S4        & S5      \\
                 & 1/ksec & 1/ksec & 1/ksec & 1/ksec & 1/ksec \\
\hline
IRS 2            & 20.4      & 27.2  	 & 47.2      & 31.3      & 36.6    \\
IRS 5N           &  0.1      &  1.0      &  0.8      &  0.8      &  0.0    \\
IRS 5            & 16.1      & 27.0  	 & 49.0      & 17.1      & 11.7    \\
IRS 6            &  1.3      &  2.4      &  2.2      &  2.7      &  2.0    \\ 
IRS 1            & 25.3      & 14.5  	 & 14.5      & 18.3      & 26.7    \\
IRS 9            &  0.5      &  2.5  	 &  0.7      &  2.3      &  0.9    \\
R CrA            &  4.7      &  6.4  	 &  5.3      & 23.7      & 15.2    \\
IRS 7W           &  1.0      &  6.1  	 &  5.1      &  1.6      &  1.9    \\
IRS 7E           &  0.7      &  0.7  	 &  0.4      &  1.3      &  1.5    \\
\hline
\label{cxoresults}
\end{tabular}

%$^1$ affected by chip gaps, taken into account via exposure maps \\
\end{table}

\subsection{Radio data}

\begin{figure}
      \includegraphics*[width=\linewidth,bb=50 170 550 670]{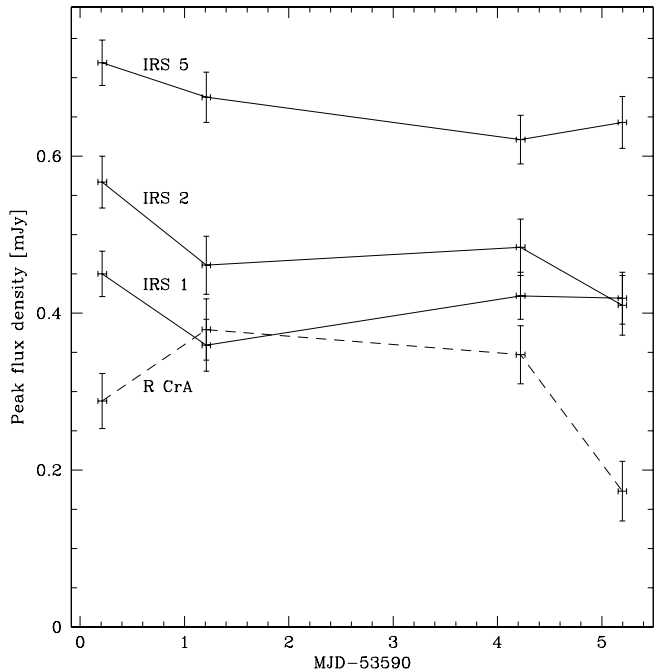}

\caption{Exemplary VLA radio flux density curves for the three class~I protostars IRS~1, 2, and 5, as well as for R~CrA. Flux densities were derived by fitting Gaussians with the AIPS task SAD. The error bars mainly take into account actual noise (1$\sigma$) rather than the quality of the fit.}
\label{rcurve}
%\end{center}
\end{figure}

The radio data show the same sources as in previous observations (Fig.~2 in \citealp{for05}), although the D array configuration of the VLA used this time results in lower angular resolution which is problematic only for IRS~7 and its immediate vicinity. The combined 2005 data is less sensitive than the combined 1998 data. No clear variability beyond an estimated uncertainty of $\sim 15\%$ in the absolute flux density scale is seen within these four epochs of observing which is why we only quote flux levels determined from the combined dataset in Table~\ref{vlaresults} while showing exemplary radio flux density curves for the three class~I protostars as well as R~CrA in Fig.~\ref{rcurve}. Also on timescales shorter than the duration of each epoch (2h), no variability was found. For comparison, the minimum and maximum flux densities from the 1998 monitoring are shown as well in Table~\ref{vlaresults}. The most striking differences are IRS~5 -- the only known class~I protostar with circularly polarized non-thermal centimetric radio emission -- showing a lower flux than before throughout this observation and R~CrA with a higher radio flux. The flux of IRS~5 is in fact so low that we cannot detect Stokes-$V$ emission which presumably is below the detection threshold, which would indeed be the case if the polarization degree was comparable to the values measured in 1998. The only X-ray flare of IRS~5 during this observing run took place during VLA maintenance time so that it could not be monitored simultaneously at radio wavelengths. R~CrA was marginally detected as a weak radio source before, now it is brighter and clearly detected. IRS~2 appears brighter than in the 1998 data, similar to levels reported before \citep{fcw98}. When comparing the radio variability, it is important to note the different timescales probed in our 1998 and 2005 monitoring campaigns. While in 1998, observations span nearly four months, in 2005 only five days are covered. Our results suggest generally lower levels of variability on these shorter timescales.

\subsection{X-ray data}

\begin{figure}
\includegraphics*[width=8.7cm, angle=-90, bb= 48 100 524 580]{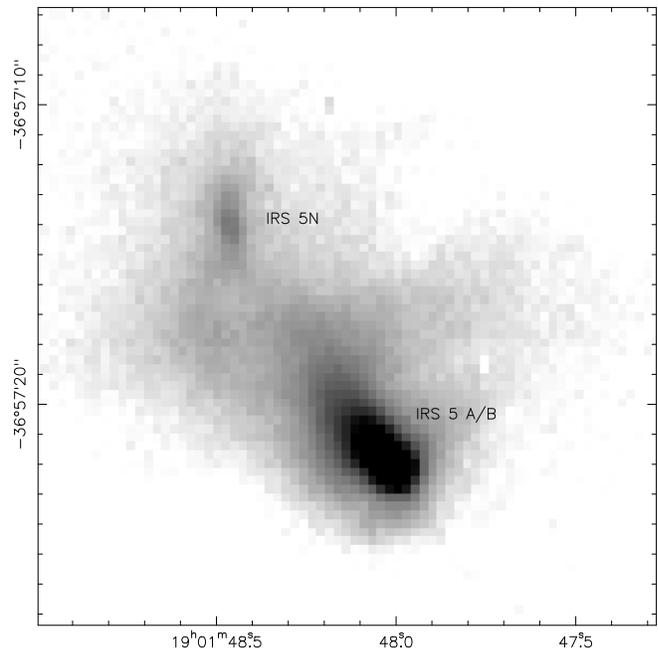}
\begin{center}
\caption{The double source IRS~5 (marginally resolved here) and its neighbouring source IRS~5N seen in $K_S$ (SofI/NTT) on August 9, 2005.}
\label{irs5sofi}
\end{center}
\end{figure}

Focusing on multi-wavelength variability here, a detailed X-ray census of the Corona Australis region is beyond the scope of this study and will appear in a separate paper \citep{fpr06}. Compared to the previous data presented by \citet{for05}, several new X-ray sources have been detected in the region of the \textsl{Coronet} cluster (Fig.~2 in \citealp{for05} vs. Fig.~\ref{vlacxo}). With the exception of the source north-east of IRS~5 (IRS~5N, detected only in S2), none of these sources is detected simultaneously at radio, near-infrared or optical wavelengths. IRS~5N was detected as a weak radio source by \citet{for05} and is also seen in the near infrared (Fig.~\ref{irs5sofi}).

Averaged count rates per observing epoch for the main \textsl{Coronet} sources are summarized in Table~\ref{cxoresults}. The count rates are similar to previous \textsl{Chandra} count rates (e.g. \citealp{for05}). The most variable sources by factors of about five are IRS~5, R~CrA, and IRS~7W. IRS~5 appears to be in a decaying flare state in epoch S3. R~CrA shows increased levels of X-ray emission in the last two epochs S4 and S5. No strong flare towards IRS~7E occured during this observing run, but the source is detected with a low and slightly variable flux level (but cf.~\citealp{ham05}).
The class~I source IRS~9 shows an X-ray flare in the second observing epoch S2. 
Both IRS~7E and IRS~9 are too weak for the analysis of meaningful light curves. Thus we only give the averaged count rates.
Confirming the detection reported by \citet{ski04}, we also see T~CrA in X-rays, south-east of IRS~7. We note that the spectral properties of the sources are similar to those reported by \citet{for05}.

There are a few interesting sources outside the VLA primary beam which merit attention in this context because they are variable and were observed in at least two wavelength regimes simultaneously: S~CrA shows variability in the first half of the observations and 2MASS~J19012872-3659317 shows a flare in the last \textsl{Chandra} observing epoch S5. The X-ray light curves are plotted in Figs.~\ref{xcurvej} and \ref{xcurvei}.

\subsection{Near-infrared and optical data}

\begin{figure*}      
\sidecaption
\includegraphics*[width=12cm,bb=20 480 540 700]{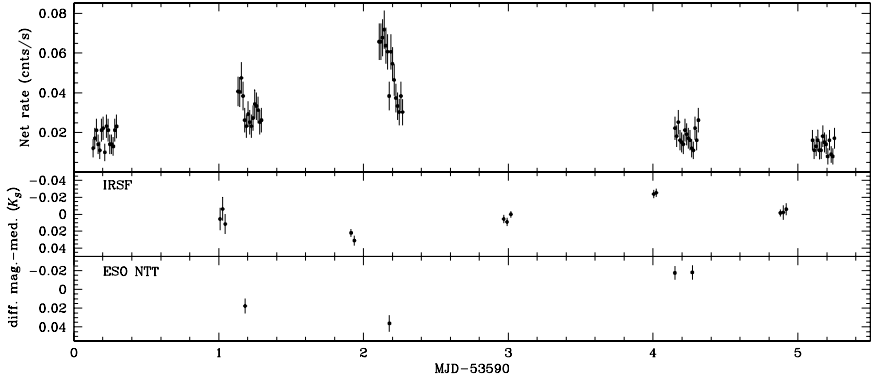}
%\begin{center}
\caption{Top panel: \textsl{Chandra} X-ray light curves of IRS~5 for the five observations S1-S5, bin size is 1~ksec. Center and bottom panel: $K_S$-band differential photometry of IRS~5 taken with SIRIUS and SofI, respectively. If anything, the $K_S$-band flux appears to slightly decrease around the enhanced X-ray activity of IRS~5. Error bars are $1\sigma$.}
\label{xcurveirs5}
%\end{center}
\end{figure*}

In a first step, near-infrared and optical data were searched for counterparts of X-ray sources. We then singled out those sources that were variable by more than 0.1~mag in the respective band. The discussion focuses on these variable sources. We note, however, that while a full discussion of all sources is beyond the scope of this paper, not all X-ray sources in this region have NIR or optical counterparts and vice versa. In general, only slow variability is seen at these wavelengths, with the notable exception of TY~CrA which showed an eclipse minimum.

Again, many sources are not covered by the VLA primary beam, but there is a good near-infrared/optical to X-ray correspondence in source coverage. Fifteen near-infrared counterparts of X-ray sources spread over the four different pointings of the SIRIUS imager were analyzed for variability in the $J$ band, sparing seven sources which were too bright in the $J$ band for analysis in the $I$ band (see below, a few of them were in fact also too bright in the $I$ band). Two class~I protostars, namely IRS~5 and IRS~9, as well as the embedded source north-east of IRS~5, are undetected in $J$ band and thus had to be analyzed in $K_S$ band.

\begin{figure*}[!ht]
\begin{minipage}{12cm}
      \centerline{a) IRS~1}
      \includegraphics*[width=12cm,bb=20 480 540 700]{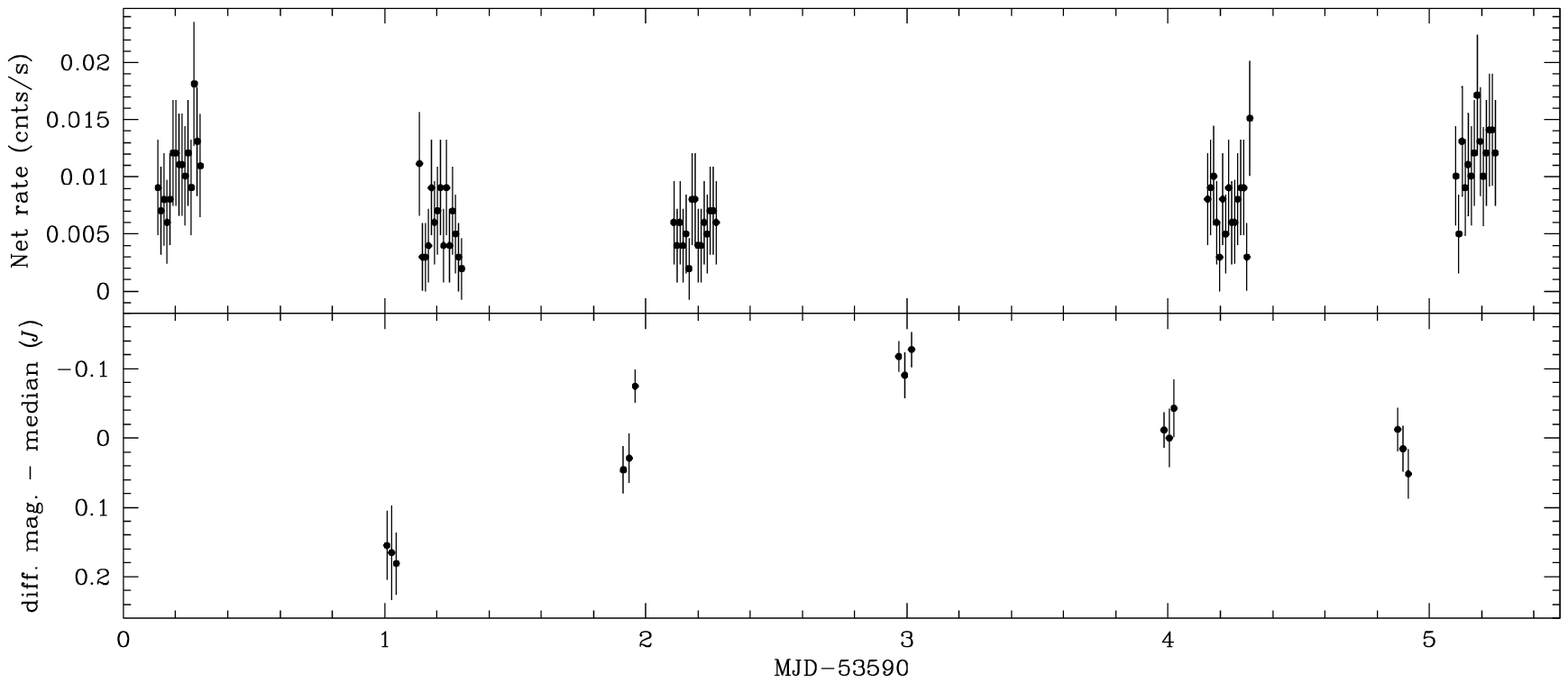}
      \centerline{b) IRS~2}
      \includegraphics*[width=12cm,bb=20 480 540 700]{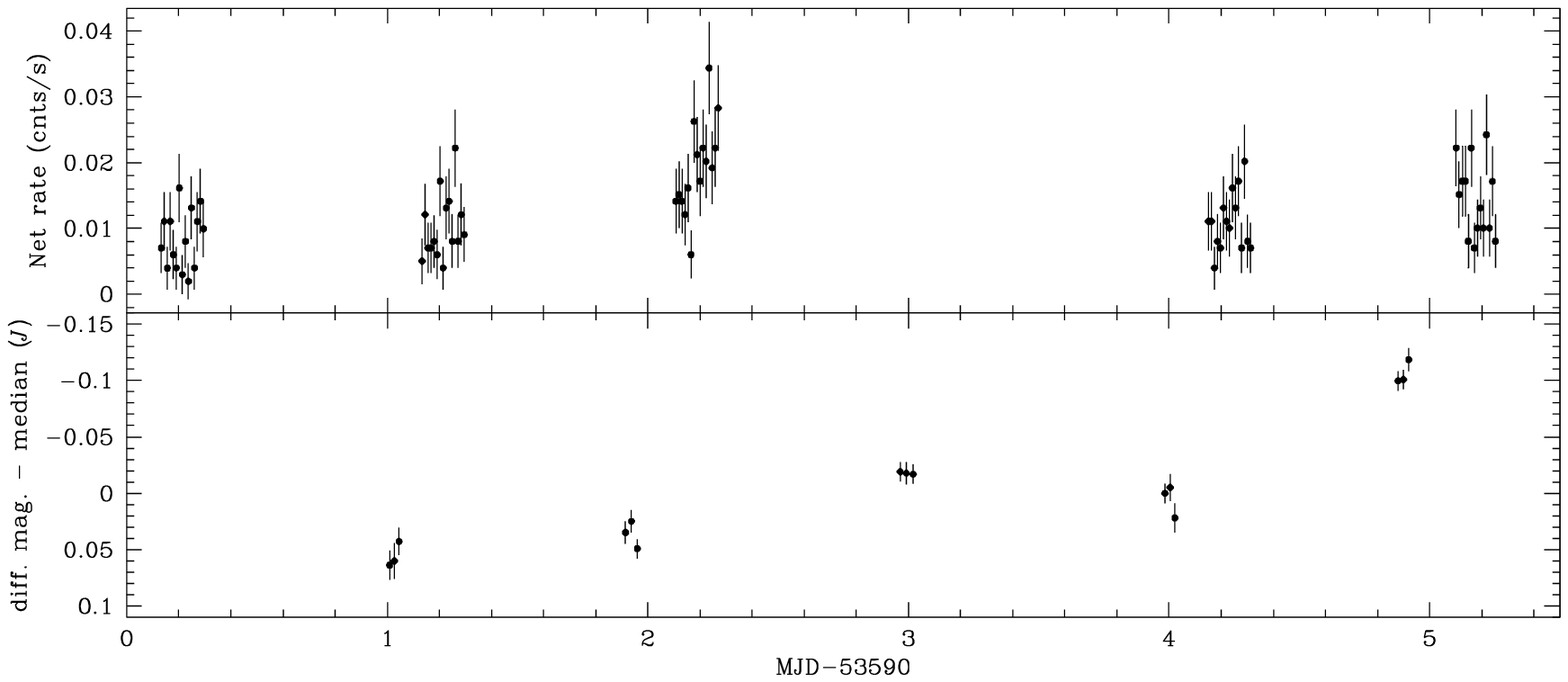}
      \centerline{c) 2MASS~J19013385-3657448}
      \includegraphics*[width=12cm,bb=20 480 540 700]{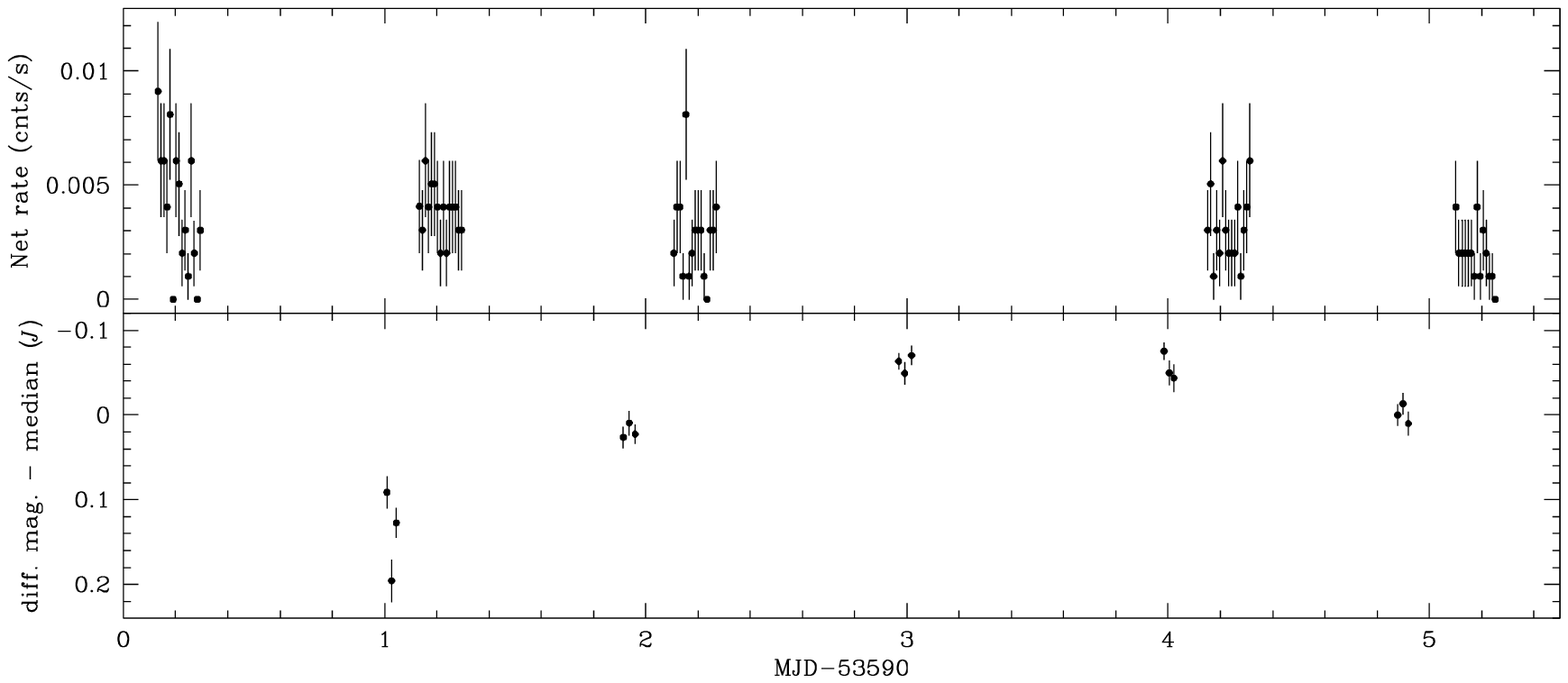}
      \centerline{d) 2MASS~J19014041-3651422}
      \includegraphics*[width=12cm,bb=20 480 540 700]{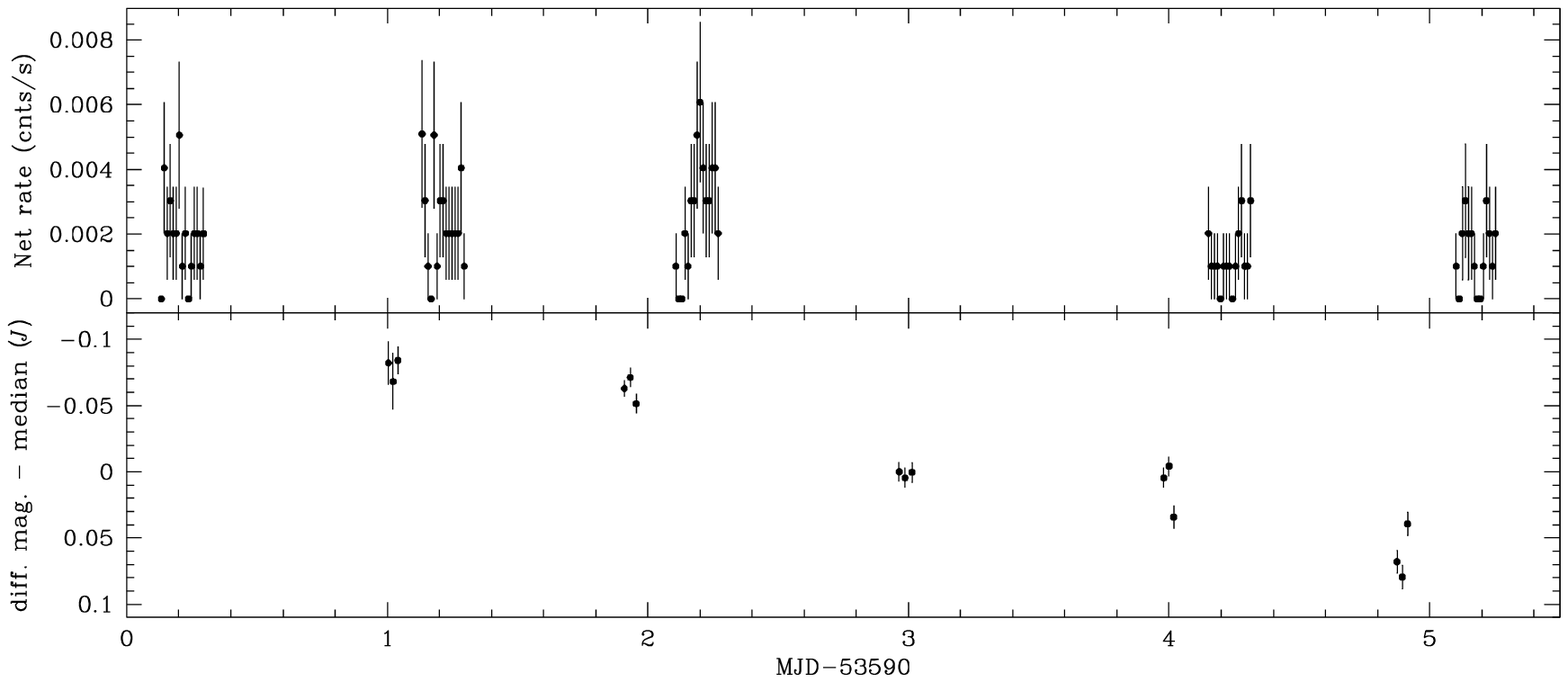}
\end{minipage}
\begin{minipage}{6cm}
\vspace{20.7cm}
\caption{\textsl{Chandra} X-ray light curves for YSOs with near-simultaneous IRSF $J$-band data in the respective lower panels (with 1$\sigma$ error bars).}
\label{xcurvej}
\end{minipage}
%\end{center}
\end{figure*}

Among the selected SIRIUS $J$-band sources, four show variability of $>0.1$~mag (see Fig.\ref{xcurvej}), namely the two class~I protostars IRS~1 and IRS~2 as well as 2MASS~J19013385-3657448, and 2MASS~J19014041-3651422. It is remarkable that this selection includes the two identified class~I protostars which are detected at all in the $J$ band. In the $K_S$ band, IRS~9, IRS~5N and IRS~5 (the latter being shown in Fig.~\ref{xcurveirs5}) vary by less than 0.1~mag, but IRS~9 and IRS~5N brightened steadily during the observations. 

%\subsection{Optical data}

\begin{figure*}[!ht]
\begin{minipage}{12cm}
      \centerline{a) R CrA}
      \includegraphics*[width=12cm,bb=20 480 540 700]{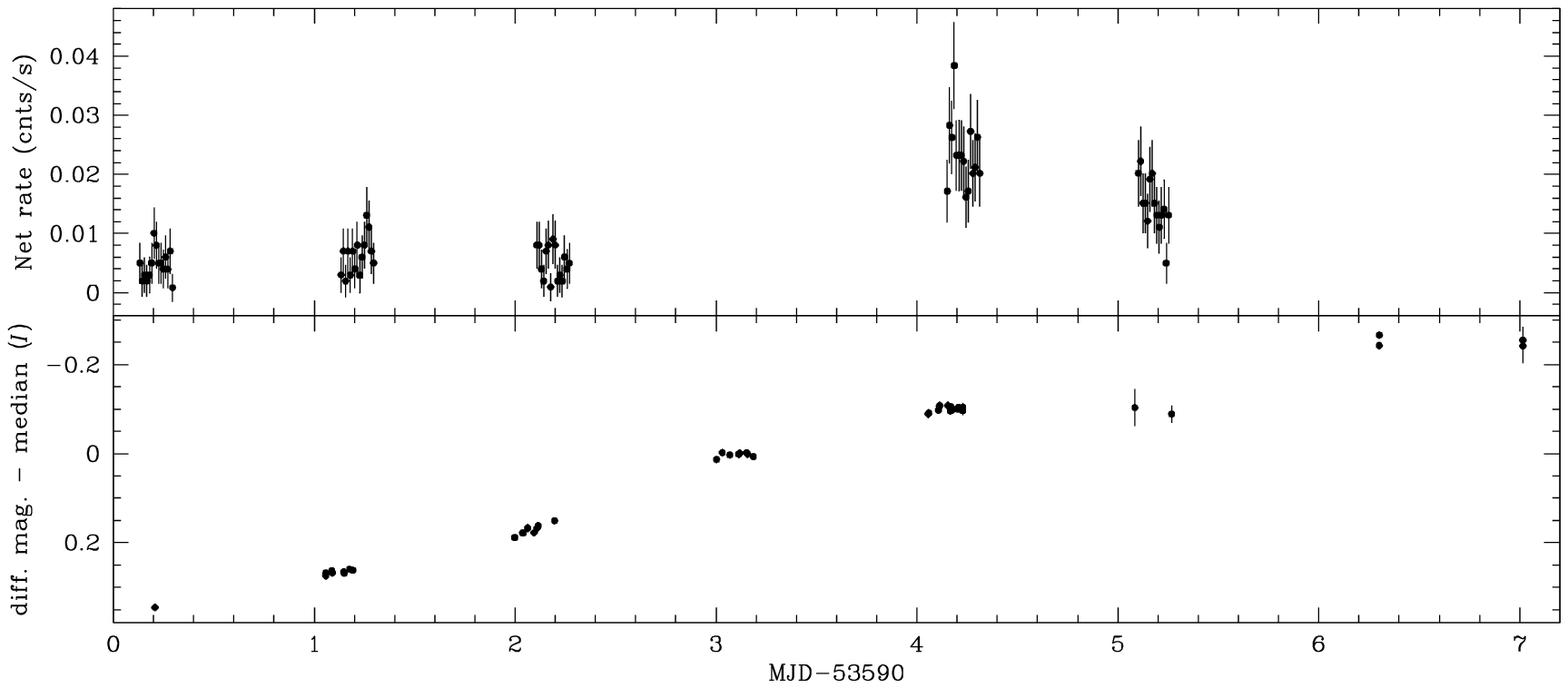}
      \centerline{b) S CrA}
      \includegraphics*[width=12cm,bb=20 480 540 700]{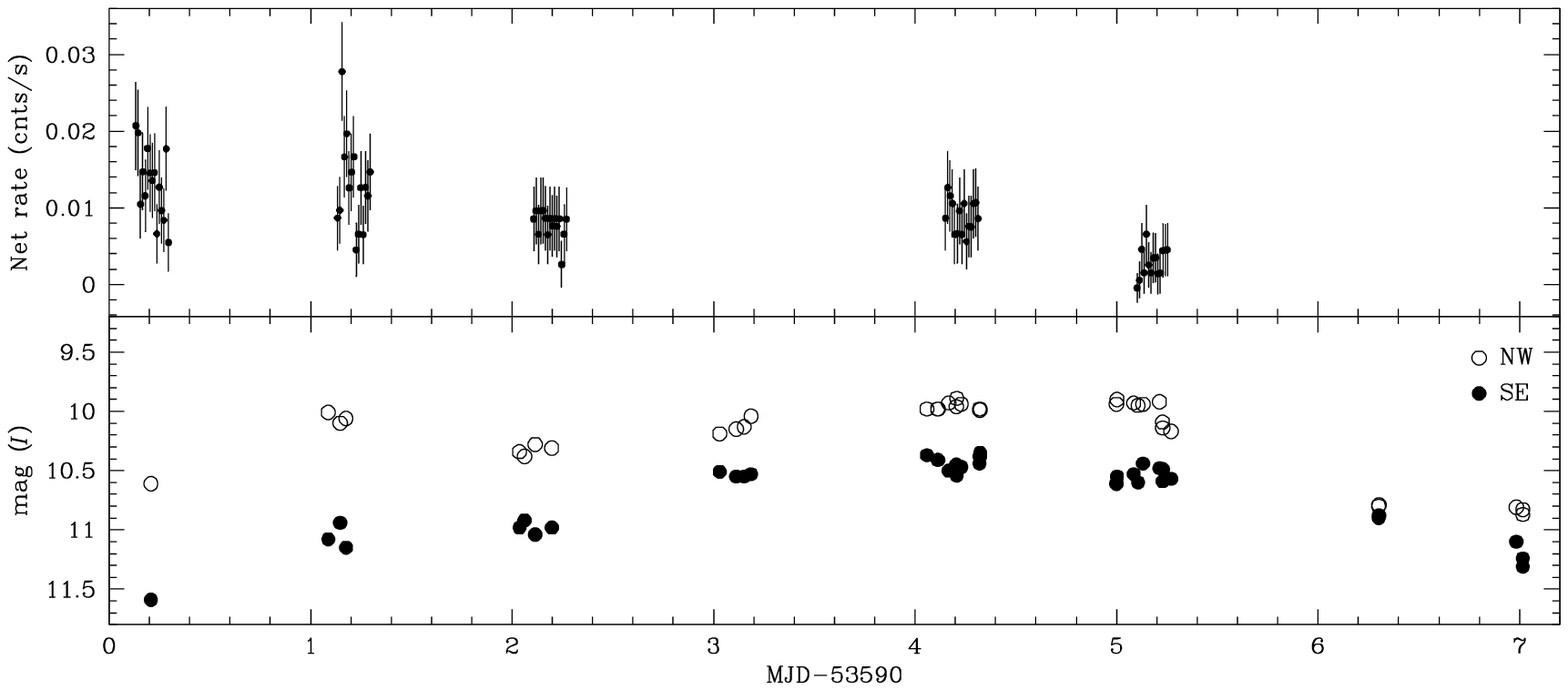}
      \centerline{c) TY CrA}
      \includegraphics*[width=12cm,bb=20 480 540 700]{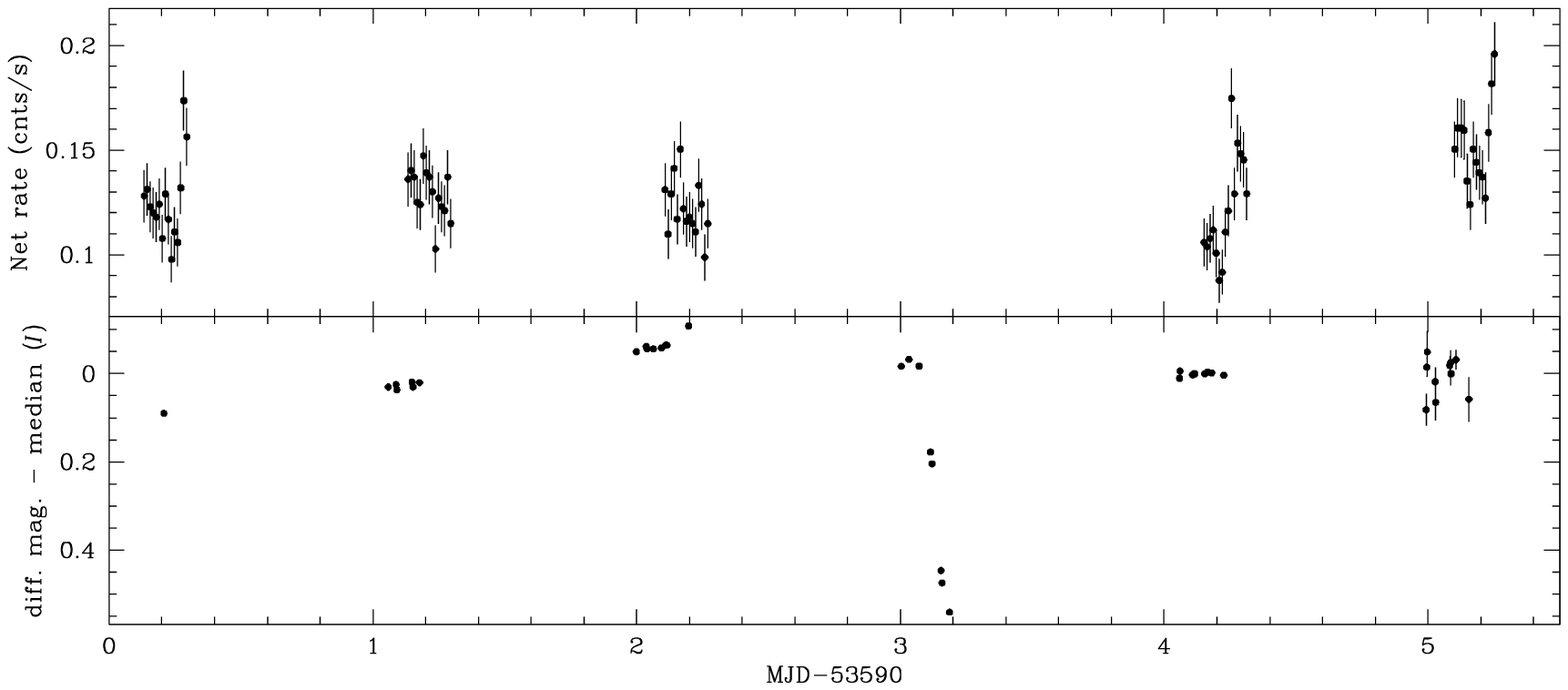}
      \centerline{d) ISO-CrA 137}
      \includegraphics*[width=12cm,bb=20 470 540 700]{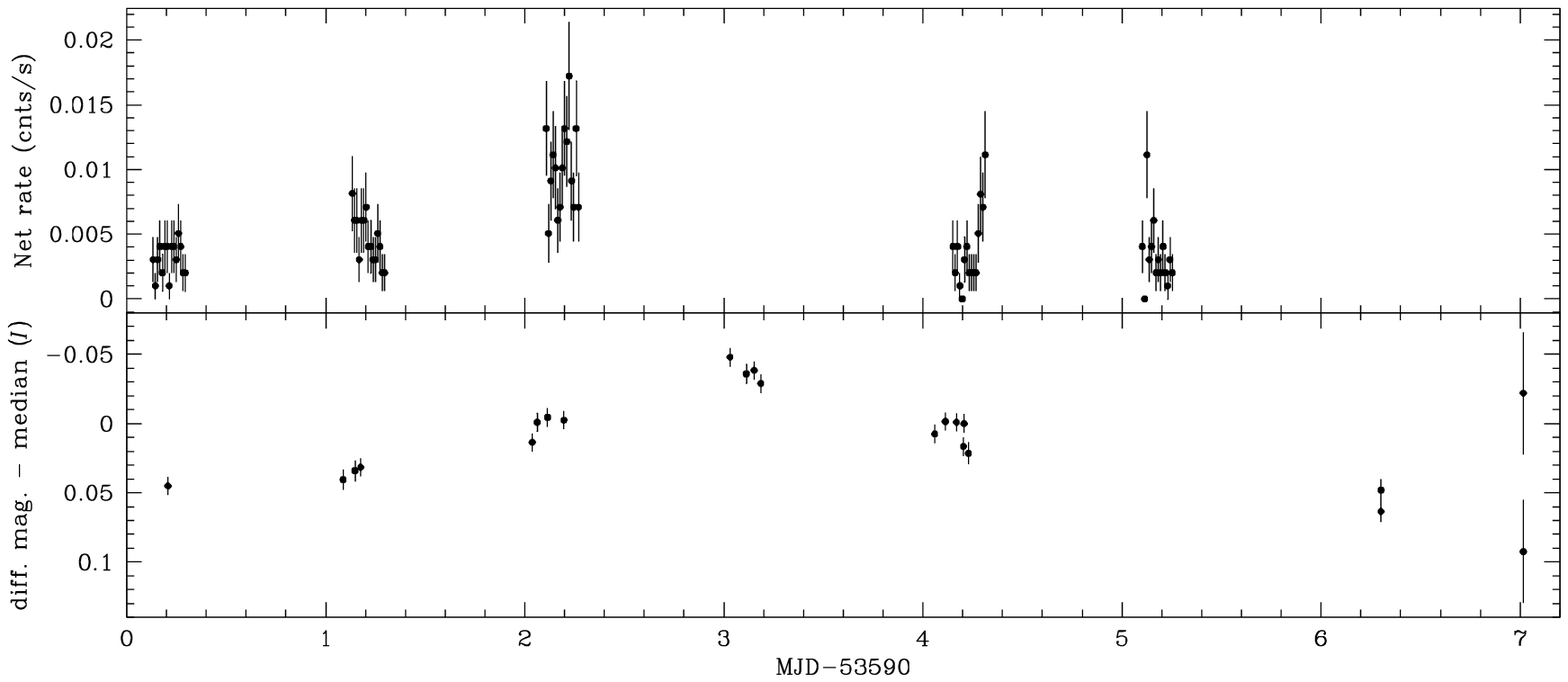}
\end{minipage}
\begin{minipage}{6cm}
\vspace{17cm}
\caption{\textsl{Chandra} X-ray light curves for YSOs with simultaneous CTIO $I$-band data in the respective lower panels (with 1$\sigma$ error bars). The $I$-band light curve for S~CrA results from a special analysis involving Gaussian fitting to its two components NW and SE (see Fig.~\ref{scrafred} and text). Figure continues on next page.}
\label{xcurvei}
\end{minipage}
%\end{center}
\end{figure*}

Most of the saturated $J$-band counterparts of X-ray sources except for HD~176386 (m$_{\rm V}$=7.30) were also detected in $I$ band and available for photometry. Several sources show variability of $>0.1$~mag, namely the Herbig Ae stars R~CrA, T~CrA, and TY~CrA, as well as the T Tauri star S~CrA and the embedded sources ISO-CrA~137 and HBC~680, an M3 dwarf WTTS \citep{neu00}. The light curves are shown in Fig.~\ref{xcurvei} and in Fig.~\ref{icurve}.

\begin{figure*}
\begin{minipage}{12cm}
      \centerline{e) 2MASS~J19012872-3659317}
      \includegraphics*[width=12cm,bb=20 470 540 700]{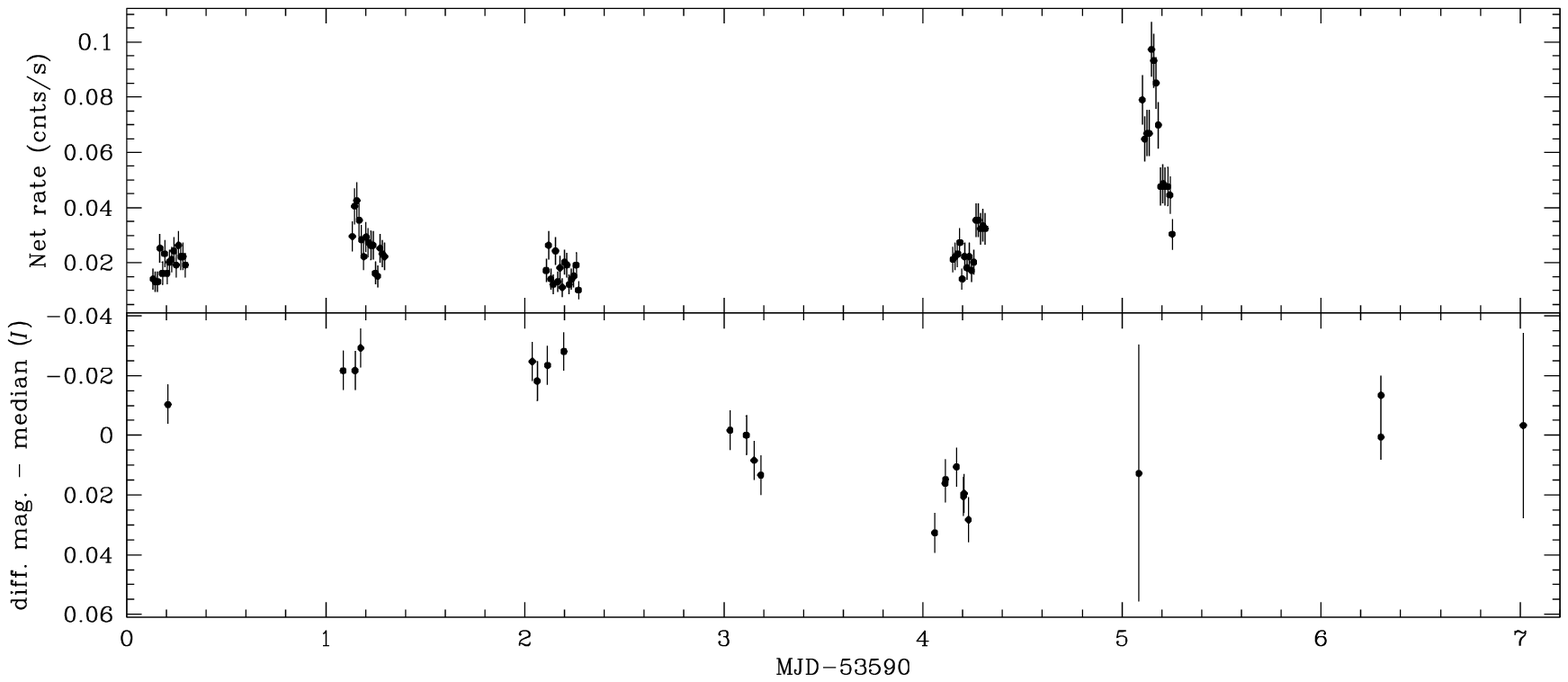}
\end{minipage}
\begin{minipage}{6cm}
\vspace{4.9cm}
{\bf Fig.~\ref{xcurvei}.} Continued.
\end{minipage}
%\end{center}
\end{figure*}

\section{Multi-wavelength variability}
\label{singlesource}

\subsection{The class~I protostars IRS~2, 5, 1, and 9}
The purpose of these observations was to, for the first time, obtain simultaneous multi-wavelength data on class I protostars. As stated above, no significant radio variability was observed during the four VLA observations. IRS~2 and IRS~1 show moderate X-ray variability (see Fig.~\ref{xcurvej}), together with uncorrelated near-infrared variability, while IRS~5 is again the most variable class~I protostar in X-rays. The enhanced activity in \textsl{Chandra} observation S3 was not covered by simultaneous radio observations. The $K_S$-band flux of IRS~5 is slightly lower before and during the S3 observation (Fig.~\ref{xcurveirs5}). The radio luminosity of IRS~5 is much lower during the observations reported here than before, with its Stokes-$V$ emission below the detection limit. IRS~5N, the weak X-ray source north-east of IRS~5 (Fig.~\ref{irs5sofi}) is detected in X-rays only in epoch S2. The source was simultaneously observed in the near-infrared with SofI/NTT, however no significant variability could be found compared to the SofI/NTT exposures taken later on except for a slight steady brightening by $<0.1$~mag throughout the SIRIUS $K_S$-band observations. While the class~I source IRS~9 shows an X-ray flare in the S2 \textsl{Chandra} observation, it remains undetected in the VLA observations as before. No significant near-infrared variability besides a slight steady brightening by $<0.1$~mag is seen in the SIRIUS $K_S$-band data (the X-ray flare was not covered simultaneously at NIR wavelengths, both light curves not shown). 

\subsection{The Herbig Ae/Be stars R~CrA, T~CrA, and TY~CrA}

The three Herbig stars R~CrA (type Ae), T~CrA (type F0e), and TY~CrA (type R8e) are among the most variable $I$-band sources in this dataset. R~CrA which appeared to be in different X-ray activity stages in the data presented by \citet{for05} is flaring continuously and is particularly active in the last two \textsl{Chandra} (S4-5) observations (Fig.~\ref{xcurvei}). The X-ray emission of R~CrA is again associated with an extremely high plasma temperature.
%\new{($>64$~keV, see Fig.~\ref{xrayspectra}).} 
Simultaneously, R~CrA continuously brightened by approximately 0.6~mag ($I$ band) throughout the observation campaign. The optical data, however, are compromised by bad weather directly after the S4 \textsl{Chandra} observation. Still it seems that the brightening stops after the onset of enhanced X-ray activity, if only to resume later. The same brightening is observed in the $U$ band (not shown). Given the earlier discussion on the origin of the X-ray emission towards R~CrA \citep{for05}, it is quite likely that two different objects are in fact observed here in optical and X-ray emission. 

T~CrA is quite variable at optical wavelengths (Fig.~\ref{icurve}), however it is only barely detected as an X-ray source. While our observing run spans several eclipse minima of the multiple Herbig Be system TY~CrA, no minimum has been observed in different bands simultaneously (see optical data in Fig.~\ref{xcurvei}).

\subsection{The classical T Tauri stars S~CrA and IRS~6}

S~CrA is a $1\,\farcs3$ visual pair, varying by up to two magnitudes and showing variable inverse P Cygni profiles in the upper Balmer lines, indicative of mass accretion. \citet{pra97} found that both components can be identified as TTSs. The source remains unresolved in our X-ray data due to its off-axis location.
Our X-ray light curve shows a quite slowly varying count rate with an apparently uncorrelated $I$-band light curve (see Fig.~\ref{xcurvei}). The shape of the optical light curve is quite complex though not unusual. 

% FRED

\begin{figure}
\includegraphics*[width=\linewidth,bb=20 270 540 670]{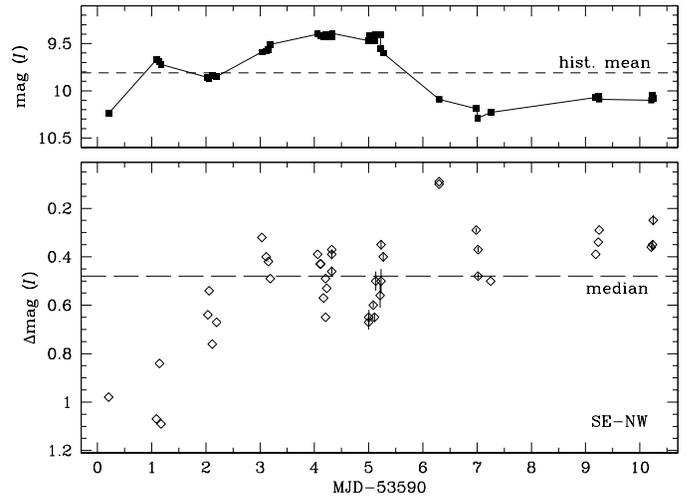}
\begin{center}
\caption{Upper panel: The total $I$-band magnitude of S~CrA, determined as described in the text. The dashed horizontal line is the historical mean in the Van Vleck Observatory data base, used as the zero point here. Lower panel: The magnitude difference between the two components of S CrA. The dashed horizontal line is the median difference.}
\label{scrafred}
\end{center}
\end{figure}

The $I$-band light curves of the NW and SE components are shown in comparison to the X-ray data Fig.~\ref{xcurvei} while Fig.~\ref{scrafred} contains the full light curve discussed here together with the magnitude difference of the two components. Both components of the binary brightened throughout the time of the X-ray observations, before drastically fading. At peak the system is about 0.3~mag brighter than the historical mean, and within 0.1~mag of the maximum brightness we have observed. There is a 0.5~mag brightening in the NW component with a duration $<$ 2 days, on the second day. The NW component is consistently brighter than the more volatile SE component. Historically (that is, in our database), the NW component is brighter than the SE component at $I$ 89$\pm$2\% of the time, with a median brightness difference of 0.45~mag. We observed a median brightness difference of 0.48~mag. The rate of change of the stars is also well within the norms. The only oddity is the slow apparently coherent variations in the two stars. We would be concerned if the stars varied exactly in phase, but 
Fig.~\ref{scrafred} shows that throughout this time the SE component is brightening relative to the NW component. Still, the concordance of the component light curves is unusual. The light travel time between the components, at a projected separation of 170~AU, is about one day. This makes it difficult to explain the rapid fading in terms of, say, a dust event in the system. 

\begin{figure}
\includegraphics*[width=\linewidth,bb=20 270 540 670]{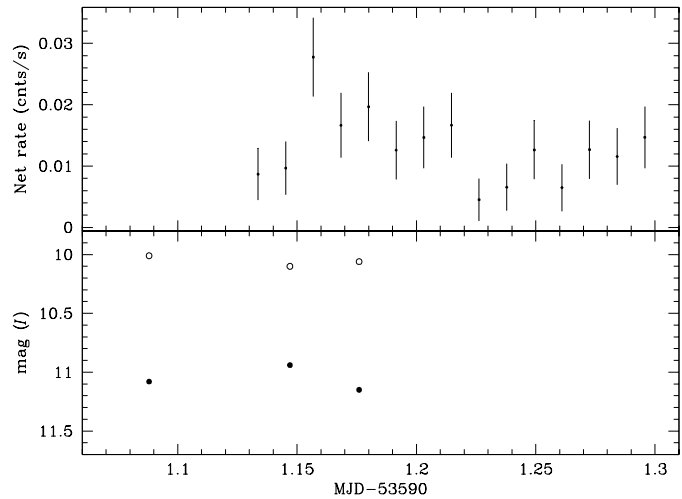}
\begin{center}
\caption{Detail of the X-ray and $I$-band light curves shown in Fig.~\ref{xcurvei}. The optical flare seen in the NW component does not appear to directly correspond to the maximum in the X-ray light curve.}
\label{scrafred1b}
\end{center}
\end{figure}

Even in this more detailed photometric analysis of the two components of S~CrA, we find no apparent correlation between the X-ray light curve and the optical light curve. While the brightest X-ray points correspond closely to the 0.5~mag flare in the NW component, there does not seem to be a direct connection (see Fig.~\ref{scrafred1b}).

% FRED^

The T Tauri star IRS~6 is detected in centimetric radio emission, at a similar level as before \citep{for05}. Low-level X-ray variability is accompanied by insignificant changes in the near-infrared emission during the observation run. 

\subsection{Additional sources}

\begin{figure}
\includegraphics*[width=\linewidth,bb=40 170 540 670]{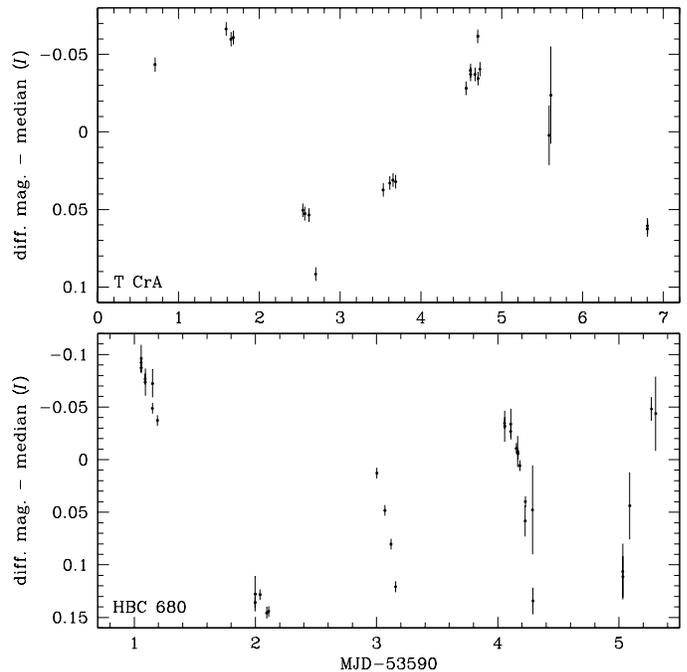}
\caption{CTIO-$I$ light curves for T CrA (upper panel) and HBC 680 (lower panel). While the former source is barely detected in X-rays, the latter remains undetected. Error bars are $1\sigma$.}
\label{icurve}
\end{figure}

Four more sources merit attention in this context because of peculiar X-ray or optical/NIR variability. The averaged X-ray count rates for these sources are listed in Table~\ref{cxoresults2}. 2MASS~J19013385-3657448 is observed at low and barely changing X-ray count rates, however shows a maximum $J$-band magnitude approximately in the middle of the observations. 2MASS~J19014041-3651422 drops continuously in $J$-band brightness, however shows fairly constant X-ray count rates. More variable X-ray emission with enhanced activity in epoch S3 is observed towards 2MASS~J19012901-3701484 = ISO-CrA 137, a source showing a well-defined $I$-band emission maximum in-between \textsl{Chandra} observations S2 and S3. While 2MASS~J19012872-3659317 shows the most pronounced X-ray variability of all these sources, the $I$-band light curve, while structured, shows only $<10\%$ (0.1~mag) variations. Only one $I$-band measurement was taken during the enhanced X-ray activity, and this data point has a large error. The flare in epoch S5 is  not covered in $U$ band due to bad weather. The M~dwarf HBC~680 is a rapidly variable source in the $I$ band (Fig.~\ref{icurve}), but was not detected in X-rays, likely due to its proximity to the edge of the field-of-view. 
 
\begin{table*}
%siehe countrates.ods
%\begin{center}
\caption[]{X-ray count rates for discussed sources outside the VLA primary beam, from the five \textsl{Chandra} observations, 0.5-10~keV, corrected for effective exposure}
\begin{tabular}{lrrrrr}
\hline
\hline
Source ID               & S1        & S2        & S3        & S4        & S5      \\
                        & 1/ksec    & 1/ksec    & 1/ksec    & 1/ksec    & 1/ksec \\
\hline
2MASS~J19012872-3659317 & 21.3      & 28.5      & 17.5      & 26.5      & 67.4\\
2MASS~J19012901-3701484 = ISO-CrA 137             &  3.8      & 6.2       & 11.8      &  4.8      &  4.1\\
2MASS~J19013385-3657448 &  3.3      & 3.0       & 2.5       &  2.7      &  1.8\\
2MASS~J19014041-3651422 &  4.0      & 4.7       & 5.2       &  2.5      &  2.9\\
\hline
\label{cxoresults2}
\end{tabular}

%$^1$ affected by chip gaps, taken into account via exposure maps \\
\end{table*}

\section{Conclusions}

YSOs in and around the \textsl{Coronet} cluster in Corona Australis, including class~I protostars, were observed simultaneously in multi-epoch X-ray, radio, near-infrared and optical observations. The main conclusions are as follows:

\begin{itemize} 
\item The sources in the \textsl{Coronet} cluster are variable in X-rays and near-infrared/optical emission, however no variability at radio wavelengths could be detected even though the measured flux densities differ from previously measured values, indicating long-term variability.
\item No clear multi-wavelength correlations in variability were found. The near-infrared variability of the class~I protostars observed here appears to be uncorrelated to their X-ray variability (with the exception of a slight -- possibly insignificant -- drop in $K_S$ band flux during enhanced X-ray activity of IRS~5). This is what would be expected in case of coronal X-ray emission -- as opposed to X-rays due to accretion -- which is also indicated by the very high plasma temperatures involved.
\item Towards the Herbig Ae star R~CrA, an apparent jump in X-ray emission is observed during continuous brightening in the optical regime. While the classical T Tauri star S~CrA shows a complex optical light curve, its X-ray emission is much less variable. Its two components were found to nearly vary synchronously in the $I$ band.
\item The radio behaviour of the polarized radio continuum source IRS~5, a class~I protostar, could not be studied due to the low flux density of the source. The source shows, however, enhanced X-ray activity in one observation.
\item Among the four sources with X-ray counterparts and an near-infrared ($J$) variability exceeding 0.1~mag, there are two class~I protostars, a disproportionate share in spite of the low numbers. In case of the $I$ band variability, six sources with variability exceeding 0.1~mag include three Herbig Ae/Be stars, one CTTS, as well as two additional embedded sources.

\end{itemize}

\begin{acknowledgements}
Thanks to Scott Wolk and \textsl{Chandra} Mission Planning for their efforts in scheduling simultaneous \textsl{Chandra} and VLA observations. 
Thanks to Angeliki Field-Pollatou for swapping observing time at CTIO, to Dipankar Maitra for carrying out part of the optical observations in service mode, and to Edgardo Cosgrove for nice observing support. Charles Bailyn on short notice kindly agreed to schedule additional simultaneous IR observations at CTIO which were then not carried out due to bad weather. Thanks to Valentin Ivanov for help in preparing the SofI observations. JF acknowledges generous support by Studienstiftung des deutschen Volkes. This work is partly based on observations collected at the European Southern Observatory, Chile (ESO Programmes 60.A-9994, 74.C-0049 and 75.C-0626) and makes use of data products of the Two Micron All Sky Survey, which is a joint project of the University of Massachusetts and the Infrared Processing and Analysis Center/California Institute of Technology, funded by the National Aeronautics and Space Administration and the National Science Foundation.
\end{acknowledgements}

\bibliographystyle{aa} % style aa.bst
\bibliography{bibmaster} % your references Yourfile.bib

\end{document}